\documentstyle[epsf]{aipproc}
\begin{document}

\title{UV Opacity in Nearby Galaxies and Application to Distant Galaxies}

\author{Daniela Calzetti}

\address{Space Telescope Science Institute}

\maketitle

\begin{abstract}
The effects of dust opacity on the radiation of nearby and distant
galaxies are reviewed. The geometrical distribution of the dust inside
the galaxy plays a fundamental role in determining the wavelength
dependence of the obscuration and the opacity of the galaxy. In the
local Universe, late Hubble type galaxies appear to contain enough
dust that corrections for the effect of obscuration become
important. This is true expecially at blue and UV wavelengths, i.e. in
the wavelength range of interest for studies of massive stars and star
formation processes. Multiwavelength observations provide a powerful
tool for characterizing the reddening caused by dust. A `recipe' is
given for removing the dust reddening and recovering the UV and
optical light in star-forming galaxies.
\end{abstract}

\section*{Opacity in the Local and Distant Universe}

Dust opacity alters the radiation from astronomical sources and the
physical quantities we derive from it. The effects of dust on the
light are twofold: a) global dimming of the radiation output from a
source and b) selective (i.e., wavelength-dependent) removal of the
radiation. Because of the wavelength dependence of the extinction, the
UV will be more affected by dust than the optical or the infrared;
sources which are optically thin in the visible (e.g., A$_V$=0.3)
become optically thick in the UV (A$_{1300~\AA}\simeq 1$). Dust,
therefore, can have a major effect on the wavelength range selected 
for studying high-mass star formation processes.  Matters are further
complicated by two additional ``characteristics'' of reddening:

$\bullet$ the details of the extinction curve are
environment-dependent. The best known cases are the diffuse ISM
extinction curves of the Milky Way (Seaton 1979), the Large Magellanic
Cloud (LMC, Fitzpatrick 1986), the the Small Magellanic
Cloud (SMC, Bouchet et al. 1985), and our twin galaxy M31 (Bianchi et
al. 1996), which are characterized by different shapes. Even within
our own Milky Way, different environments have different extinction
curves (Cardelli, Clayton \& Mathis 1989).

$\bullet$ In extended objects (HII regions, galaxies, etc.) the
`effective' obscuration is a function of the geometrical distribution
of the dust relative to the emitters (e.g., Witt, Thronson \& Capuano
1992). An extinction curve can be defined or used in a straightforward
manner only when all the dust is foreground to the emitter, as in the
case of stars. In galaxies, the dust is usually mixed in a complicate
fashion with the dust; geometry becomes the dominant factor in
determining the wavelength dependence of the obscuration (Natta \&
Panagia 1984, Calzetti et al. 1994, CKS94 hereafter).  Even in the
simplest (unrealistic) case that all the dust is in a shell
surrounding the galaxy, back-scattering from the farthest regions of
the dust shell into the line of sight produces an effective
obscuration which is greyer than the `standard' extinction curves.

While it is generally agreed that galaxies, at least the late Hubble
types, contain dust, less agreement exists on how effective the dust
is in obscuring the emerging light (e.g., the Cardiff Meeting, Davies
\& Burstein 1995). What makes dust elusive in galaxies is the lack of
obvious emission/absorption features, the potential confusion between
dust reddening and the ageing of the stellar population, and the grey
`net' obscuration often produced by the combination of dust scattering
and geometry (Witt et al. 1992). Because of these reasons, indirect
methods must be usually employed to determine the opacity of a
galaxy. Multiwavelength observations to characterize selective
obscuration (Peletier et al. 1994, Bosma et al. 1992, CKS94, Calzetti
1997, C97 hereafter), variations of the galaxy opacity with
inclination (Giovanelli et al. 1994, Burstein, Willick \& Courteau
1995), and extinction of background sources by the foreground galaxy
(White, Keel, \& Conselice 1996, Berlind et al. 1997, Zaritski 1994,
Gonzalez et al. 1997) are among the most common methods.

The central regions and the arms of spiral galaxies are likely to be
opaque, with A$_B\approx$1 or larger, while the interarm regions are
generally transparent, with A$_B\approx$0.3 (White et al. 1996,
Giovanelli et al. 1994, Berlind et al. 1997). Dust may be present in
the haloes of galaxies (A$_B\approx$0.1, Zaritsky 1994). Active star
formation (SF) is usually associated with strong far-infrared emission
from the dust heated by the massive stars (Rieke et al. 1980, Soifer
et al. 1987, Helou 1986). The effects of dust in star-forming regions
are the result of two competitive processes; on the one side, massive
stars are born in the dusty environments of molecular clouds; on the
other side, an evolving stellar population tends to blow away or
destroy the surrounding dust through supernovae explosions and massive
star winds.

Evidence exists for presence of dust at high redshifts. Damped
Ly-$\alpha$ Systems (DLAs) around z$\simeq$2--3 have metallicities
around 1/10--1/15 solar and dust/gas around 3--20\% of the Milky Way
value (Pei, Fall \& Bechtold 1991, Pettini et al. 1997a). The DLAs may
not be representative of the high redshift galaxy population as a
whole, but if so (e.g., Wolfe 1995, Pettini et al. 1997b), the metal
abundances at z=3 are not primordial, and dust, which comes with
metals, may be a concern (see, however, Pettini \& Bowen 1997). In a
Universe which is 1/6--1/4 its present age, only a relatively small
fraction of gas has been locked up in stars; gas column densities are
larger than in the Local Universe and even a small dust/gas ratio can
imply a measurable reddening. In the redshift range z$\sim$2.5--3.5,
the rest-frame UV is shifted to optical wavelengths; for instance, a
spectrum in the wavelength range 4,000--10,000~\AA~ corresponds to the
rest-frame range 1,000-2,500~\AA~ of a z=3 galaxy. Therefore, standard
ground-based observational techniques are sensitive to the rest-frame
UV emission from distant galaxies, namely to a wavelength region
potentially heavily impacted by dust reddening.

\section*{A ``Recipe'' for Reddening}

Despite the complications discussed in the previous section, dust
reddening at UV and optical wavelengths can be ``treated'', at least
in galaxies and galaxy regions where massive stars dominate the
radiation output (CKS94, Kinney et al. 1993, 1994, Calzetti et
al. 1996, C97). This includes a wide range of
extragalactic objects, from the centrally concentrated starbursts in
spirals to the Blue Compact Dwarfs. In regions of SF, the massive star
population responsible for the nebular line emission is also
responsible for most of the UV radiation. The spectral shape of the UV
emission ($>$1200~\AA) is relatively constant over a relatively large
range of ages, because we are observing the Rayleigh-Jeans part of the
massive stars' spectrum; the non-ionizing photons which make the UV
spectrum are less age-sensitive than the ionizing photons (i.e.,
nebular line emission); the latter disappear before appreciable
changes in the UV spectral shape can be observed. If the UV spectrum
is fit as F($\lambda$)$\propto\lambda^{\beta}$, the UV index $\beta$
has values between $-2.5$ and $-2$ for a reddening-free, ionizing star
population (Leitherer \& Heckman 1995, LH95 hereafter). The relation
between UV stellar continuum and ionized gas emission has proven
crucial for pinning down the selective effects of dust obscuration in
star-forming galaxies. Various diagnostics have been constructed from
multiwavelength data (Figure~1a and 1b; CKS94 and C97), extending the
relation between stellar continuum and nebular emission from the UV to
the K band. 

%
\def\putplot#1#2#3#4#5#6#7{\begin{centering} \leavevmode
\vbox to#2{\rule{0pt}{#2}}
\includegraphics{#1}
\end{centering}}
%

Adopting the standard notation:
\begin{equation}
F_{obs}(\lambda) = F_0(\lambda)\ 10^{-0.4 E_s(B-V)\ k(\lambda)},
\end{equation}
with F$_{obs}$($\lambda$) and F$_0$($\lambda$) the observed and 
intrinsic fluxes, respectively, the selective attenuation of the 
stellar continuum k($\lambda$), normalized to k$(B)-$k$(V)=1$,  
can be expressed as:
\begin{eqnarray}
k(\lambda) &=& [(1.86-0.48/\lambda)/\lambda - 0.1]/\lambda + 1.73 \ \ \ \ \ \ \ 0.63\ \mu m \le \lambda \le 1.0\ \mu m \nonumber \\
           &=& 2.656\, (-2.156 + 1.509/\lambda - 0.198/\lambda^2 + 0.011/\lambda^3) + 4.88\nonumber \\
           & &\ \ \ \ \ \ \ \ \ \ \ \ \ \ \ \ \ \ \ \ \ \ \ \ \ \ \ \ \ \ \ \ \ \ \ \ \ \ \ \ \ \ \ \ \ \ \ \ \ \ \ 0.12\ \mu m \le \lambda < 0.63\ \mu m.
\end{eqnarray}

\putplot{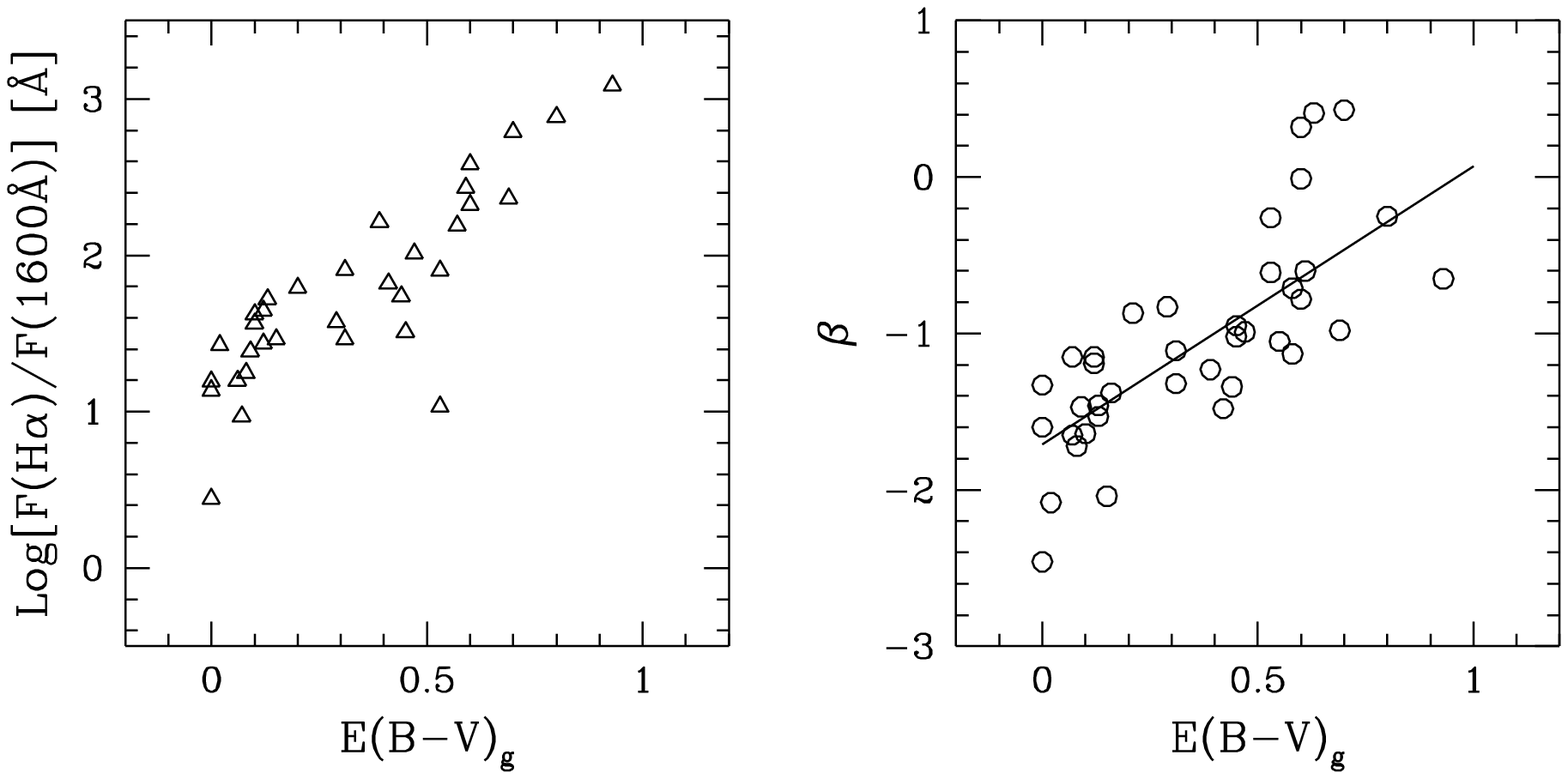}{3.0 in}{0}{80}{80}{-50}{-340}
\vskip0.1in
\footnotesize\noindent
{\bf FIGURE 1.} {\bf a) (left panel).} The ratio of H$\alpha$ to UV emission, 
observed in a sample of star-forming galaxies, is shown as a
function of the color excess of the ionized gas E(B$-$V)$_g$ 
(measured from the Balmer line ratio H$\alpha$/H$\beta$). The UV is
centered at 1600~\AA. If both nebular lines and UV continuum radiation
are due to massive stars, the correlation is explainable as an effect
of selective reddening at different wavelengths. {\bf b) (right
panel)} The UV index $\beta$ is plotted as a function of E(B$-$V)$_g$
for the same sample of galaxies.  The best linear fit is shown as a
continuous line. In a stellar population, the values of $\beta$ are
relatively constant for the range of ages where the nebular hydrogen
lines (H$\alpha$, H$\beta$, etc.) are detectable; therefore, the
observed correlation between $\beta$ and E(B$-$V) is attributable to
dust reddening, and not ageing, with the UV spectrum becoming redder for
increasing values of the color excess.
\normalsize

\vspace*{0.2 in}
The selective attenuation is shown in Figure~2a in comparison with two
extinction curves, the Milky Way and the SMC. The comparison is purely
illustrative and should not be taken at face value, because the dust
attenuation of galaxies is conceptually different from
the dust extinction of stars. The latter measures strictly the dimming
effect of the dust between the observer and the star, while the former
folds in one espression (equation~2) a variety of effects: extinction,
scattering, and the geometrical distribution of the dust relative to
the emitters. One comparison is, however, licit: the 2175~\AA~bump,
which is a prominent feature of the Milky Way extinction curve, is
absent in the attenuation curve. Gordon et al. (1997, see, also, this
Conference) have proven that the absence of the feature cannot be
explained either with scattering or dust geometry, and must be {\it
intrinsic to the extinction curve} of the ISM in the star-forming
galaxies.

Expressions~(1) and (2) can be used to derive the intrinsic spectral
energy distribution F$_0$($\lambda$) of the star-forming region, once
the effective color excess E(B$-$V)$_s$ of the stellar continuum is
known. Because of the geometrical information folded into the
expression of k($\lambda$), E(B$-$V)$_s$ is not a straighforward
measure of the total amount of dust between the observer and the
source (as in the case of individual stars). The relation between
E(B$-$V)$_s$ and the color excess E(B$-$V)$_g$ of the ionized gas is:
\begin{equation}
E(B-V)_s = 0.44 E(B-V)_g.
\end{equation}
Here, the color excess of the ionized gas is derived from the Balmer
decrement (or any suitable set of atomic hydrogen emission lines) and
the application of a `standard' extinction curve. The selective
extinction of the Milky Way, LMC or SMC curves has similar values at
optical wavelengths (Fitzpatrick 1986), so any of these curves can be
used for the ionized gas. In addition, a foreground dust distribution
appears to work well for the gas when moderate extinctions,
E(B$-$V)$_g\approx$0.1--1, are present. 

Regions of active SF may be inhospitable to dust; supernovae
explosions and massive star winds generate shock waves and, possibly,
mass outflows (Heckman et al. 1990). Shocks and outflows likely
destroy or remove the dust from inside the region; only the external
(foreground) dust survives (Calzetti et al. 1996), accounting for the
observed gas reddening geometry. This simple interpretation does not
account, however, for Equation~(3): stars are on average a factor 2
less reddened than the ionized gas (Fanelli et al. 1988, CKS94). The
factor 2 difference in reddening implies that the covering factor of
the dust is larger for the gas than for the stars (C97). Indeed, while
the nebular emission requires the presence of the ionizing stars, the
UV and optical stellar continuum is contributed also by non-ionizing
stars. Ionizing stars are short-lived and remain relatively close to
their (dusty) place of birth during their entire lifetime, while the
long-lived non-ionizing stars have time to `diffuse' into regions of
lower dust density. If this is the case, stars and gas will not occupy
the same regions (Calzetti et al. 1997), and stellar continuum and
nebular emission should be largely uncorrelated. Why then does the
reddening of the stellar continuum correlate with the reddening of the
ionized gas, as implied by Figure~1? For both the correlation and
Equation~(3) to be valid, the ageing and diffusion of the stars must be
compensated by the production of new massive stars. In other words,
the SF event must have a finite duration and cannot be
instantaneous. A lower limit to the SF duration can be placed by
remembering that the crossing time of a region of $\sim$500~pc is
about 50~Myr for a star with v=10~km/s.

Whichever the interpretation, expressions~(2) and (3) are purely
empirical results, and are independent of any assumption on the
geometry of the dust distribution and on the details of the dust
extinction curve. They yield probably the most appropriate reddening
corrections for the integrated light of extended star-forming regions
(or galaxies, see Figure~2b).

\putplot{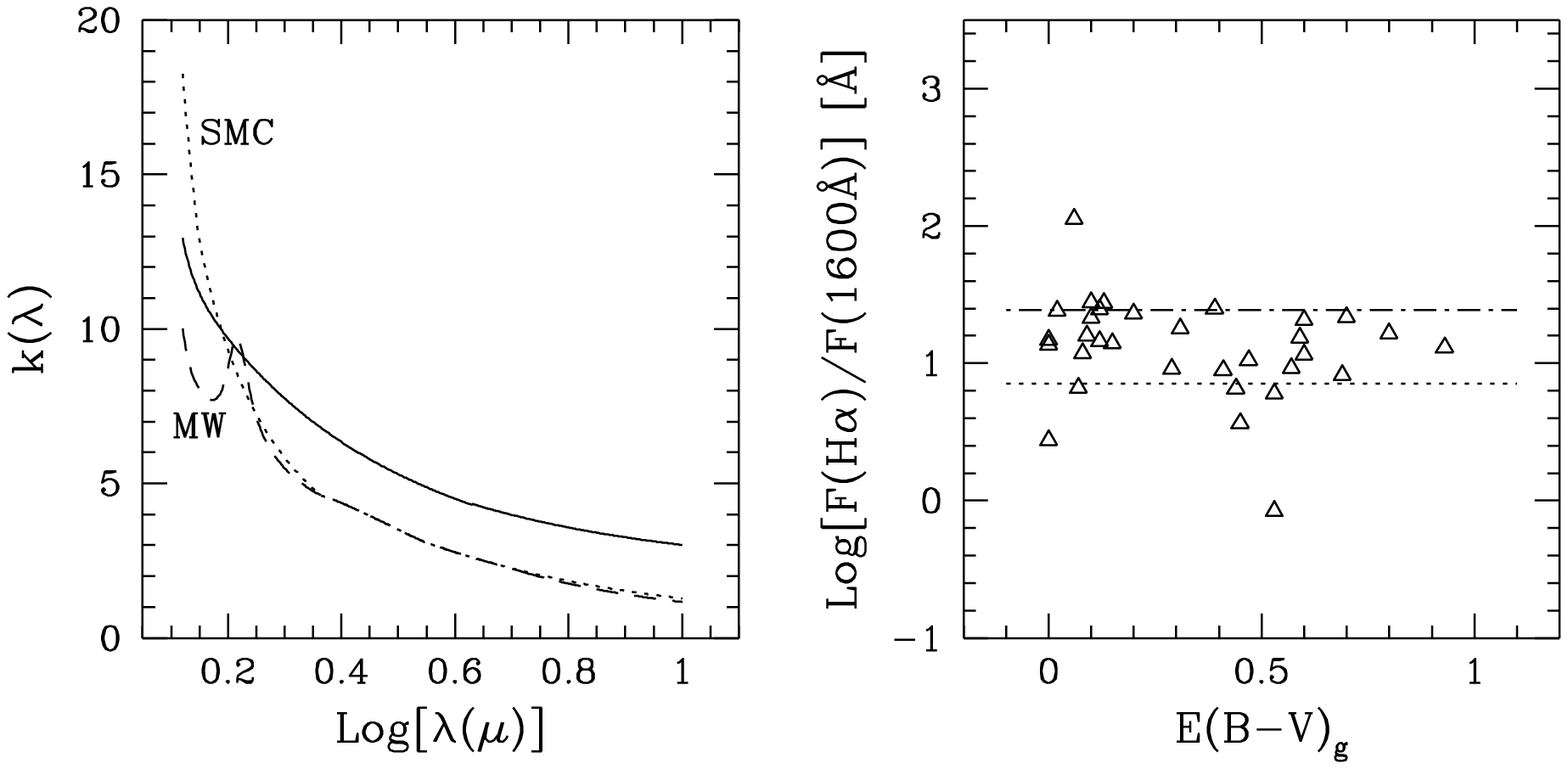}{3.0 in}{0}{80}{80}{-50}{-340}
\vskip0.1in
\footnotesize\noindent
{\bf FIGURE 2.} {\bf a) (left panel).} The attenuation curve of Eq.(2)
(continuous line) is compared with the diffuse ISM extinction curves
of the Milky Way and the SMC (dashed and dotted line,
respectively). The attenuation is normalized to k$(B)-$k$(V)=1$. The
most prominent feature of the Milky Way curve, the 2175~\AA~bump, is
absent in the attenuation curve. {\bf b) (right panel)} The
H$\alpha$/UV emission for the same galaxies of Figure~1a is shown
after correcting the UV emission for dust attenuation using Eqs.(2)
and (3) and E(B$-$V)$_g$ from the Balmer decrement. The H$\alpha$
emission is corrected using the Balmer decrement and the Milky Way
extinction curve. The two horizontal lines represent the range of
expected values for H$\alpha$/UV for stellar populations undergoing
continuous star formation in the two extreme cases of
0.1--100~M$_{\odot}$ and 0.1--30~M$_{\odot}$ Salpeter Initial Mass 
Function (top and bottom line, respectively). Eq.(2) does recover the UV 
emission expected for the observed H$\alpha$ emission.
\normalsize

\section*{Distant Galaxies}

The recently discovered galaxy population at z$\sim$3 (Steidel et
al. 1996) and the UV-dropouts in the Hubble Deep Field (Williams et
al. 1996) have re-opened the debate on the SF and metal production
histories of the high-redshift Universe (see Madau et al.  1996). One
of the current issues is to understand which fraction of the total
high-redshift SF these galaxies represent (e.g., Pettini, this
Conference, Dickinson, 1997; Madau, 1997). Here, the potential impact
of dust reddening on the estimates of the high-redshift SF and metal
production rates are briefly discussed (see, also, Meurer et al. 1997,
and, for a detailed analysis, Dickinson 1997).

By the nature of the detection criterion (Steidel \& Hamilton 1993),
all the z$\simeq$3 galaxies have intense UV emission, tracer of recent
massive SF (Steidel et al. 1996, Giavalisco et al. 1996). From the
spatial extension of the UV emission (a few kpc, Giavalisco et
al. 1996), the galaxies have been forming massive stars for at least
100 Myr and probably more; this is the amount of time required by the
`SF wave' to cross a typical galaxy scale, 1~kpc, if it travels at the
sound speed, v$\approx$10~km/s.  The spectrum of a dust-free galaxy
which is forming stars at a constant rate since T$_d$=100~Myr has
$\beta\sim-2.45$, and, if T$_d$=1~Gyr, $\beta\sim-2$, with little
dependence on the metallicity down to 1/10 solar (LH95, Bruzual \&
Charlot 1996, C97). However, the UV indices measured in the high-z
galaxies are redder than these values; the observed {\it average} UV
index of the distant galaxies is $\beta\simeq-1.1$ (Meurer et
al. 1997). Taking into account corrections for the Lyman Forest
absorption, the galaxies UV spectral energy distributions become 
slightly bluer, $\beta\simeq -1.2$ to $-1.5$ (cf. Dickinson 1997), but
still too red to be compatible with on-going constant-rate (or slowly
decreasing) SF. To reconcile the observed {\it average} UV index with
the expectation for a star-forming galaxy, three possible scenarios
(or a combination thereof) can be invoked: 1) the massive end of the
Initial Mass Function of the z=3 galaxies is steeper than the
present-day IMF; 2) the z=3 galaxies are `aged star-forming galaxies':
the SF happened in the past and massive stars are no longer
being formed in large quantities; ageing reddens the UV index of a
stellar population; 3) the UV emission of the z=3 galaxies is reddened
by dust. Ageing and reddening are particularly important, because both
scenarios suggest that the SF and metal production rates inferred from
the observed UV emission of the high-z galaxies are underestimates of
the global SF and metal production rates.

$\bullet$ {\it The IMF.} There is increasing evidence that in local
starbursts the slope and high-mass cut-off of the IMF above
5--10~M$_{\odot}$ is independent of the environment (Massey et
al. 1995, Stasi\'nska \& Leitherer 1996). Invoking a different IMF
between the high redshift and the present-day galaxies thus requires
an explanation of why the IMF has changed with time. In this regard,
it should be remembered that the metallicity of high-z Damped
Ly-$\alpha$ Systems is not drastically lower than the values observed
today in Blue Compact Dwarf galaxies. The shape of the IMF is thus
unlikely to have changed from z$\sim$3 to the present, though the issue
is still open.

$\bullet$ {\it Ageing.} The average high-z galaxy is assumed to be a
dust-free ageing stellar population. Let's assume the stellar
population was generated in a constant-SFR, star-forming episode of
duration T$_d$. After T$_d$ years, the SF stops, and the stellar
population is left ageing for T$_a$ years; the ``average'' galaxy is
observed at T$_a$ to account for the red UV index. For T$_d$=100~Myr,
a time T$_a\simeq$150~Myr must lapse before $\beta\simeq -1.2$. At
this stage, the UV flux around 1500~A is about f=55 times lower than
the value at t$\le$100~Myr. A T$_d$=1~Gyr star-forming population
would need T$_a\simeq$20--50~Myr to reach $\beta\simeq -1.2$.  In this
case, the decrease in UV flux relative to the peak is about a factor
f$\sim$4--6.  If the observed galaxies are located at z=3~(3.5), the
peak of SF (t$\le$T$_d$) would occur at z$>$3.12~(3.7) in an open
Universe (H$_o$=50~Mpc/km/s and q$_o$=0.0) and at z$>$3.25~(3.9) in a
flat Universe (q$_o$=0.5). This peak has not been observed (e.g.,
Madau et al. 1996), although the incompleteness of the samples may
play a large role here; of course, the redshift of the peak can be
pushed to higher values if the SF doesn't interrupt abruptly after
T$_d$ (as in our simplified model), but is a decreasing function of
time.

$\bullet$ {\it Reddening.} If the {\it average} z=3 galaxy can be
described by a 1~Gyr old stellar population undergoing SF at a
constant rate, an effective color excess E(B$-$V)$_s\simeq$0.15 is
needed to change the UV index from $-2$ to $-1.2$. This modest color
excess implies a UV attenuation A$_{(1600~\AA)}\simeq$1.65. The
intrinsic UV emission, and corresponding SFR, is then underestimated
on {\it average} by a factor f$\sim$5. Any stellar population
younger than 1~Gyr will have bluer intrinsic UV spectra, and will
imply larger reddening corrections (cf. Meurer et al. 1997).

A simple argument can be used to infer that the amount of metals
produced in the z=3 galaxies by the observed SF event are comparable
for the ageing and the reddening scenarios. The injection of metals
into the ISM is proportional to the total number of massive stars
produced. If the IMF is the same in the two scenarios, the number of
massive stars generated by the SF event is proportional to the
observed SFR, sfr$_{obs}$, to the factor $f$ by which sfr$_{obs}$
underestimates the peak SFR, and to the duration of the burst,
T$_d$. The quantity sfr$_{obs}$ comes from the observed UV emission
and is the same in both scenarios. In the ageing case, $f\times
T_d=$55$\times$100~Myr and 4$\times$1~Gyr, respectively; in the
reddening case, $f\times T_d=$5$\times$1~Gyr. Therefore, both
scenarios predict the same amount of metals produced at z$\sim$3--4
over the lifetime of the SF event. More sophisticated models for the
stellar populations and evolution of the high-z galaxies would
probably still give numbers in this ballpark. Metal production thus
does not appear to be a good discriminant between ageing and reddening
in the high-z galaxies.

Both reddening and ageing predict that the observed UV emission from
the distant galaxies underestimates the peak SFR; there is an
important difference between the two, though: in the first case, the
underestimated quantity is the `current SFR' at z$\sim$3; in the
second case, the quantity is the `recent past SFR' a redshift beyond
3.  Probably the most direct way to prove whether the high-z {\it
average} UV spectra are red because of dust obscuration or ageing will
be to measure the intensity of the hydrogen Balmer lines: these are
more directly related to the number of ionizing photons than the UV
emission, and can help constraining the average age of the massive
star population.

\section*{Conclusions}

We have seen that, despite the instrinsic complexity of the dust
opacity in external galaxies, in certain cases the problem of dust
obscuration is treatable. A `recipe', under the name of `attenuation
curve', is available for the reddening correction of the integrated
light from star-forming regions. The strength of the starburst
attenuation curve is that its derivation is purely empirical and does
not rely on models. The curve is therefore applicable at least to the
class of objects it has been derived from: the central star-forming
regions of galaxies.

One of the open problems is understanding the limits of applicability
of the curve. In the case of isolated HII regions, where the SF
processes are less energetic than in the case of starbursts, the dust
can survive the less harsh environment and be uniformly distributed
with the stars. For this reason, the attenuation curve will not be
generally applicable to HII regions. Similar arguments can be used in
the case of `quiescent' galaxies.

Other open problems are the meaning of Equation~(3), namely, the
discrepancy in global attenuation between stars and gas though they
are still correlated, and the physical meaning, if any, of the
effective color excess E(B$-$V)$_s$ of the stellar continuum. These
and other issues will require further investigation.

\vspace*{0.3cm}
The author acknowledges valuable discussions with Tim Heckman, Max
Pettini, Gerhardt Meurer, Mauro Giavalisco, and Claus Leitherer.

\end{document}